\documentclass[12pt,english]{article}
\usepackage{amstex,amssymb,amscd}
\begin{document}
\hoffset=-8mm
\voffset=-8mm

\title{Thermodynamics of a finite system of classical particles with
short and long range interactions and nuclear fragmentation\\}

\medskip
\author{J. Richert$^{\rm a}$, P. Wagner$^{\rm b}$, M. Henkel$^{\rm c}$,
J.M. Carmona$^{\rm d}$\\}

\maketitle
\thispagestyle{empty}
\bigskip\noindent
$^{\rm a}$ Laboratoire de Physique Th\'eorique,
Universit\'e Louis Pasteur\\
3, rue de l'Universit\'e, 67084 Strasbourg Cedex, France\\ \\
$^{\rm b}$ Institut de Recherches Subatomiques, IN2P3-CNRS and 
Universit\'e Louis Pasteur\\
BP28, 67037 Strasbourg Cedex 2, France\footnote{Unit\'e
Mixte de Recherche CNRS No 7500}\\ \\
$^{\rm c}$ Laboratoire de Physique des Mat\'eriaux,
Universit\'e Henri Poincar\'e Nancy I\\ 
BP239, 54506 Vand\oe uvre les Nancy Cedex, France\footnote{Unit\'e
Mixte de Recherche CNRS No 7556}\\ \\
$^{\rm d}$ Departamento de F\'{\i}sica Te\'orica, Facultad de Ciencias\\
Universidad de Zaragoza, 50009 Zaragoza, Spain\\

\vspace{5mm}
\begin{abstract}

\hspace{-5mm}
We describe a finite inhomogeneous three dimensional system of classical
particles which interact through short and (or) long range interactions
by means of a simple analytic spin model. The thermodynamic properties of
the system are worked out in the framework of the grand canonical ensemble.
It is shown that the system experiences a phase transition at fixed average 
density in the thermodynamic limit. The phase diagram and the caloric curve
are constructed and compared with numerical simulations. The implications
of our results concerning the caloric curve are discussed in connection
with the interpretation of corresponding experimental data.
\end{abstract}

\medskip
 
PACS : 25.70.Pq, 64.60.Cn, 64.60.Fr\

\medskip

{\it Key words} : Nuclear fragmentation. Phase transitions. Spherical
model. Metropolis Monte Carlo simulations.
 
\newpage
\setcounter{page}{1}
\section{Introduction}

Up to now, our knowledge about the properties of highly excited nuclei generated
by means of energetic particle or heavy ion collisions is still rather limited.
In practice, the only well-established information concerns universal scaling
properties~\cite{A457} of fragment size multiplicities which are observed in the
final stage of the reaction. These observables can be reproduced by means of
simple, classical microscopic models. 
They can be interpreted as a sign for the existence of a phase
transition of percolation type.

One major handicap towards further understanding of excited nuclear matter is
concerned with the distribution of the available energy among the nucleons
in the excited system. At present it is not possible to say whether the
highly excited expanding system can be considered to be in some thermal
equilibrium or not. We know from specific experimental results that the nucleus 
is able to pick up and distribute huge amounts of energy, more or less equally,
over internal degrees of freedom, at least in specific experimental
circumstances. Theoretical interpretations in terms of so called dynamic
models~\cite{J.Aichelin} show that this is however not necessarily fully established.

If one nevertheless assumes that equilibrium is reached, 
the central point of interest
concerns the thermodynamic properties of finite and infinite nuclear matter up
to very high energies and its connection with the universality properties of
fragment distributions. During the last years several interesting attempts have been
performed in this direction on the theoretical side. They rest on generic models, the
Ising~\cite{SK443}, the lattice gas and related models [4-7]
and mean field approaches~\cite{jnde}. They constitute
interesting attempts towards a unified description of the global thermodynamics
and fragment generation in nuclear systems.

The work which we develop here takes place in the same framework.
In practice we work out a classical, fully analytic realistic 3-dimensional model
in the framework of the grand canonical ensemble. The analytic treatment allows for
a clear insight into the correlations between the microscopic structure and the global
thermodynamic properties of finite systems and their thermodynamic limit. It is able
to incorporate both short and (or) long range interactions. Hence it can be used to take
the Coulomb interaction into account and to investigate the thermodynamic properties of 
other classical microscopic systems. Finally many-body correlation functions which
allow to define fragment sizes can be worked out analytically.

In the sequel we shall essentially develop the expression of the partition function
and its properties, and work out the relations between energy, temperature and
density of the system. Indications about the determination of correlation functions
will be given at the end of the letter and developed in the future.

\section{The model: grand canonical partition function for a system with short
range interactions}

We consider a 3-dimensional system of classical particles confined in a volume
$V=(Nd)^3$ where $N^3$ is the number of cubic cells of linear dimension $d$ (the
extension to a parallelepipedic volume $V=N_1N_2N_3d^3$ is straightforward). Each
cell $k$ is occupied by a particle $(s_k=1)$ or empty $(s_k=0)$. Nearest neighbour
particles interact by means of an attractive 2-body potential of strength 
$-V_0\;(V_0>0)$. The Hamiltonian of the system reads
\begin{equation}
H_0=\sum_{k=1}^A p_k^2/2m-V_0\sum_{<n\cdot n\cdot>_1}s_is_j
\end{equation}
where $A$ is the total number of particles and $<n\cdot n\cdot>_1$ stands for
``nearest-neighbour sites $(i,j)$" in each space direction.

We shall show that
this is the inhomogeneous counterpart of the homogeneous analytic model which we
developped and discussed in ref.~\cite{boose}.
The model (1) can be brought into a form which looks more closely like the
Ising model by introducing the spin variable
\begin{equation}
\sigma_k=2s_k-1
\end{equation}
Up to the kinetic term, $H_0$ then becomes an Ising Hamiltonian in a magnetic field
\begin{eqnarray}
H_0=\sum_{k=1}^A p_k^2/2m&-&(V_0/4)[\sum_{<n\cdot n>_1}\sigma_k\sigma_\ell+
6\sum_{k=1}^{N^3}\sigma_k+3N^3]
\end{eqnarray}
The conservation of the number of particles $A$ is expressed through the constraint
\begin{equation}
(1/2)\sum_{k=1}^{N^3}(1+\sigma_k)=A
\end{equation}
In order to get an exactly solvable model we replace the Ising-like model~(3)
by the spherical model~\cite{berlin}. It has been known since a long time that the
spherical model is exactly solvable in any number of dimensions, in the presence of
symmetry-breaking fields, for very general (including long-range) interactions and
even in the presence of competing interactions or disorder~[9-17].

Yet it does display, in more than two and less than four dimensions, a second order 
phase transition with non-trivial (i.e. non mean field) critical exponents. The
spherical model represents therefore an ideal test bench on which complicated concepts
can be checked analytically.
  
To introduce it, we replace the Ising spins $\sigma_k=\pm1$ in~(3) by continuous real
variables $\sigma_k\in[-\infty,\,+\infty]$ and constrain these variables through
the relation
\begin{equation}
\sum_{k=1}^{N^3}\sigma_k^2=N^3
\end{equation}
Practically it is easier to implement the constraints (4), (5) only in the mean
through Lagrange multipliers $\lambda$ and $\mu$. In the thermodynamic limit
$N\to\infty$ it is known that this leads to the same result~\cite{lewis}. The mean
spherical Hamiltonian can then be written as
\begin{equation}
H=H_0+\lambda\sum_{k=1}^{N^3}\sigma_k^2+(\mu/2)(\sum_{k=1}^{N^3}\sigma_k+N^3)
\end{equation}
and the partition function becomes
\begin{equation}
Z=\int_{-\infty}^{+\infty}\prod_{k,i}dp_k^i\int_{-\infty}^{+\infty}\prod_k
d\sigma_k\exp(-\beta H)  .
\end{equation}
We remark that going over from the Ising model to the spherical model
implies the replacement
\begin{equation*}
\prod_{k=1}^{N^3}\sum_{\{\sigma_k=\pm1\}}\;\to\;\prod_{k=1}^{N^3}
\int_{-\infty}^{+\infty}d\sigma_k   .
\end{equation*}
The quantity $p_k^i$ is the momentum of particle $k$ in the space direction
$i$ $(i=1,\ldots,n)$. In the sequel we fix $n=3$ and assume periodic boundary
conditions in the 3 space directions.

The constraints 
\begin{eqnarray}
-\beta^{-1}\,\partial\ln Z/\partial\lambda&=&N^3\nonumber\\
-\beta^{-1}\,\partial\ln Z/\partial\mu&=&A
\end{eqnarray}
fix the parameters $\lambda$ and $\mu$. This completes the definition of the model.

At first sight the spherical constraint may appear bizarre, introducing an effective
long-range interaction between the spins $\sigma_k$. But this is only apparent and can be
interpreted as follows. Consider an $m$-component spin variable $S_a$, $a=1,\ldots,m$
of unit length $\sum\limits_{a=1}^mS_aS_a=1.$ The $O(m)$ vector model with Hamiltonian
$H_{O(m)}=-J\sum\limits_{<n\cdot n>_1}\sum\limits_{a=1}^mS_i^aS_j^a$ can be shown to
converge to the spherical model in the limit $m\to\infty$~\cite{stanley}
(for $m=1,2,3$ one recovers the familiar Ising, $XY $ and Heisenberg models
respectively).

In our case, the procedure of replacing the model (1) by the spherical
model (6-8) leads to an extension which describes the fact that the particles
do not lie at constant distance from each other, since their centres in cells of
volume $d^3$ can be found at random inside each occupied cell. This is in fact
the essence of an earlier modelisation of the system~\cite{elattari,boose}. 
Allowing $s_k$ (resp. $\sigma_k$) to vary continuously can be interpreted
in practice as the description of a system for which
$\sigma_k=\pm1$ and an interaction $V_{<n\cdot n>_1}$ which takes values on the
interval $[0,+\infty]$. But events
with large value of $V_{<n\cdot n>_1}$ are strongly suppressed 
because the largest part of their contributions are generated by $\sigma'_k$s
lying in a strip of values, due to the quadratic form of $H$ in (7) and the
constraint (5) on the volume. Our 
interpretation does not take care of the fact that the positions of a set of interacting
particles are correlated and hence, so are the two-body strengths $V_{<n\cdot n>_1}$.
However, we expect that this is not a major problem because of the fact that our
calculations concern the partition function $Z$ which delivers average quantities and
hence these correlations should not show a strong effect.

In order to diagonalise the free energy in the exponent of $Z$ we Fourier-transform
$\{\sigma_k\}$ to $\{\tau_j\}$. In each space direction
$$\tau_\ell=N^{-1/2}\sum_{k=1}^Ne^{-i\ell k} \sigma_k    .  $$
Then
\begin{eqnarray}
&&\sum_k\sigma_k\sigma_{k+1}=\sum_\ell e^{-i\ell}\tau_\ell\tau_{-\ell}
\nonumber\\
&&\nonumber\\
&&\sum_k\sigma_k^2=\sum_\ell \tau_\ell\tau_{-\ell}
\end{eqnarray}
The quadratic form in $\{\tau_\ell\}$ can be explicitly integrated and
\begin{eqnarray}
&&Z(\beta,A,N^3)\propto(2\pi m)^{3A/2}\cdot \beta^{-(3A+N^3)/2} 
\cdot \exp(-\beta\mu N^3/2+3\beta V_0N^3/4)\nonumber\\
&&\phantom{Z(\beta,A,N^3)\propto(2\pi m)^{3A/2}}\cdot 
\exp\left\{\beta\left[(\mu-3V_0)/2\right]^2N^3/\left[4(\lambda-3V_0/2)\right]\right\}\nonumber\\
&&\phantom{Z(\beta,A,N^3)\propto(2\pi m)^{3A/2}\cdot}
\cdot 
\prod_{\{i_1,i_2,i_3\}}\left[\lambda-V_0(\cos\xi_{i_1}+\cos\xi_{i_2}+
\cos\xi_{i_3})/2\right]^{-1/2}
\end{eqnarray}
with $\xi_{i_k}=2\pi i_k/N\quad (i_k=0,1,\ldots,N-1)$ and the product runs over
all possible values of $(i_1, i_2, i_3)$. The first term on the r.h.s. of~(10)
corresponds to the kinetic energy contribution.

We define
\begin{eqnarray*} 
M&=&-[1+2(\beta N^3)^{-1}\partial\ln Z/\partial\mu]=-(\mu-3V_0)/[4(\lambda-3V_0/2)]
\nonumber\\
&=&-1+2A/N^3
\end{eqnarray*}
which is constant and related to the density through 
$$M=2\rho/\rho_0-1$$
where $\rho_0$ is the normal density of nuclear matter. Then equations (8) 
and (10) lead to
\begin{eqnarray}
&&\sum_{i=0}^{N^3-1}[\lambda(\beta)-V_0\alpha_i/2]^{-1}=2N^3\beta(1-M^2)\nonumber\\
\rm with&&\alpha_i\equiv\alpha_{i_1i_2i_3}=\sum_{k=1}^3\cos\xi_{i_k}\nonumber\\
\rm and&&\mu(\beta)=4[3V_0/2-\lambda(\beta)]M+3V_0   .
\end{eqnarray}
Equation (11) is a kind of dispersion relation with $N^3$ poles. We look for the
solution which corresponds to $\lambda>V_0\alpha_{\max}/2$
 which is positive $(\alpha_{\max}=3)$
and guarantees that the integrals in the expression of $Z$ make sense.
 
From (11) one can immediately see that in the thermodynamic limit $(N\to\infty)$
\hfill\break
$\lambda(\beta)\to V_0\alpha_{\max}/2$. This shows a singularity in the total free
energy, even though the free energy per particle remains finite.

For $M=0\;(\rho/\rho_0=1/2)$, this transition is continuous and occurs
at the critical temperature~\cite{singh}
\begin{eqnarray*} 
&&\beta _c=2W_3(0)/V_0\\
\rm where&&W_3(0)=(1/2)\int_0^\infty e^{-3x}I_0^3(x)dx=0.2527
\end{eqnarray*}
$I_0(x)$ being a modified Bessel function~\cite{abramowitz}.

It is now possible to construct the caloric curve for fixed density. The energy per
particle is given by
\begin{eqnarray} 
\epsilon&=&{3\over 2\beta}-{V_0\over A}{\partial\ln Z\over\partial \theta} 
={3\over 2\beta}-V_0\left\{{3N^3\over4A}[2M(M+1)+1]\right.\nonumber\\
&&+{1\over4A}\sum_{i=0}^{N^3-1}\left.{\alpha_i
\over\left(\lambda\beta-\theta\alpha_i/2\right)}\right\}
\end{eqnarray}
with $\theta=\beta V_0$.
The first term in the expression corresponds to the kinetic energy contribution.

Contrary to the model introduced in ref~\cite{singh} one sees that the solutions
$(\lambda,\mu)$ obtained through relations (11) are of the same type for any value
of $M$ because here $M$ is a constant quantity. Hence one concludes that the 
phase transition is of the same order for any fixed density.

The energy $\epsilon$ given by (12) is a continuous function of $\beta$. The
specific heat $C_V$ reads
\begin{eqnarray}
C_V&=&-\beta^2(d\epsilon/d\beta)=3/2-(V_0/4A)\bigg\{\sum_{i=0}^{N^3-1}
{\alpha_i\over(\lambda-V_0\alpha_i/2)}\nonumber\\
&&+\beta(d\lambda/d\beta)\sum_{i=0}^{N^3-1}
{\alpha_i\over(\lambda-V_0\alpha_i/2)^2}\bigg\}
\end{eqnarray}
This quantity is again continuous in $\beta$. It is in fact well
known~\cite{itzykson,berlin} 
that the system shows a phase transition at $\beta_c(M)$ where
$C_V\sim(\beta_c-\beta)^{-\alpha}$ and $\alpha=-1$. There appears however a
discontinuity in the derivative of $C_V$ with respect to $\beta$ in the
thermodynamic limit. This can indeed be guessed numerically in Fig.~1 
for increasing sizes of the system. We also remark that the convergence 
towards the $N \to \infty$ limit appears to be relatively rapid, even
around the critical point. For $\beta>\beta_c(M)$ the specific heat is
constant.

\section{Extension of the description to a system with short and (or) long 
range interactions}

The model can be easily extended to the case where particles interact by means of
short and (or) long  range two-body interactions. Hence one can for instance 
investigate the effects of the Coulomb interaction in the excited nuclear system.

The Hamiltonian can be written in the general form 
\begin{eqnarray}
H&=&K+V=K-[V_{01}\sum_{<n\cdot n.>_1}s_is_j+V_{02}\sum_{<n\cdot n.>_2}s_is_j +
\cdots\nonumber\\
&&\cdots+V_{0n}\sum_{<n\cdot n>_n}s_is_j]+\mu\sum_is_i+\lambda\sum_i(2s_i-1)^2
\end{eqnarray}
where $K$ is the kinetic energy and $<n\cdot n.>_k$ stands for 
$k^{\rm th}$ nearest neighbour in all space directions and
\begin{eqnarray*}
V_{01}&=&V_S+V_\ell^{(1)}\\
V_{02}&=&V_\ell^{(2)}\\
\vdots&&\\
V_{0n}&=&V_\ell^{(n)}
\end{eqnarray*}
with $V_S>0$ and $V_\ell^{(i)}<0$ $(i=1,\ldots,n)$.  $V_S$ stands for the short
range potential and $V_\ell^{(k)}$ for the long (Coulomb) contribution. We parametrise
this strength in the following way
\begin{equation}
V_\ell^{(k)}=(Z_p/A)e^2/kd
\end{equation}
where $Z_p$ is the total number of protons. Here again we introduce periodic boundary
conditions on the variables $\sigma_i=2s_i-1$.

The present procedure smears out the charge over all existing particles. Hence the 
interaction is not taken into account exactly, but in the average
over the occupied parts of the volume, since protons are
not explicitly distinguished from neutrons. This should not affect strongly the
calculations since one expects that the charges are effectively distributed 
approximately uniformly in the ratio of $Z_p/A$ over the set of occupied cells. It
should also be mentioned that, as in the case of the short range interaction, the
particles interact only along the three space directions and not the directions of
the diagonals. These contributions can be approximately taken into account by a
consequent renormalization of the interaction strength.

The grand canonical partition function can now be worked out by following the same
procedure as before. Notice that the extension of the spin variables $\{\sigma_k\}$
to the continuum limit introduces, as previously in section~2, 
values of $\vert V_{0k}\vert$, in the interval $[0,+\infty]$. If one follows 
the interpretation developed there, the long range interaction in (13) can 
in principle take large values. However, like there, large values of $V_{0k}$
are suppressed because of the quadratic form of $H$ and the constraint~(5).
After some algebra one gets
\begin{eqnarray}
&&Z(\beta,A,N^3)\propto(2\pi m)^{3A/2}\cdot\beta^{-(3A+N^3)/2}
\cdot\exp(-\beta\mu N^3/2+3\beta W_0N^3/4)\nonumber\\
&&\phantom{Z(\beta,A,N^3)\propto(2\pi m)^{3A/2}}
\cdot\exp\left\{\beta\left[(\mu-3W_0)/2\right]^2N^3/4(\lambda-3W_0/2)\right\}
\nonumber\\
&&\phantom{Z(\beta,A,N^3)\propto(2\pi m)^{3A/2}}
\cdot\prod_{\{i_1,i_2,i_3\}}\left[\lambda-\sum_{p=1}^nV_{0p}(\cos\xi_{i_1}^{(p)}
+\cos\xi_{i_2}^{(p)}+\cos\xi_{i_3}^{(p)})/2\right]^{-1/2}
\end{eqnarray}
with $\xi_{i_\ell}^{(p)}={2\pi\over N} pi_\ell\quad (i_\ell=0,1,\ldots,N-1)$, 
$(p=1,\ldots,n)$ and $W_0=\sum\limits_{p=1}^nV_{0p}$

\vspace{5mm} 
In the same way as before, $\lambda(\beta)$ and $\mu(\beta)$ are obtained through
\begin{eqnarray}
&&\sum_{\{i_1,i_2,i_3\}}\left[\lambda(\beta)-\sum_{p=1}^nV_{0p}(\cos\xi_{i_1}^{(p)}
+\cos\xi_{i_2}^{(p)}+\cos\xi_{i_3}^{(p)})/2\right]^{-1}
=2N^3\beta(1-M^2)\nonumber\\
&&M=-1+2A/N^3\nonumber\\
\rm and&&\mu(\beta)=4(3W_0/2-\lambda(\beta))M+3W_0
\end{eqnarray}
The energy per particle $\epsilon(\beta)$ is the same as the one given by (12)
with $V_0$ replaced by $W_0$,
\begin{equation*}
\epsilon={3\over2\beta}-{1\over A}\cdot\sum_{p=1}^nV_{0p}
{\partial\ln Z\over\partial \theta_p}\qquad{\rm with}\qquad
\theta_p=\beta V_{0p}
\end{equation*}
and an expression similar to (13) for the specific heat $C_V$.

\section{Results}

The typical behaviour of the caloric curve for different densities is shown 
in Figs.~2-4 for the case of an attractive short range, an attractive long range
and a combination of an attractive short range with a repulsive long range
interaction respectively. In each case, the energy increases with respect 
to the temperature and one does not observe the presence of a plateau, even though
the description predicts a phase transition at every density. One observes simply
some more or less pronounced kink in the slope of the curve which corresponds in
practice to the temperature at which the transition occurs. This is observed for all
types of 2-body interactions. As already mentioned in section~3,
the reason for this smooth behaviour is related to
the fact that in the spherical model the specific heat remains finite at
the transition point even at the thermodynamic limit. Only the derivative
of the specific heat shows a discontinuity at the transition points. From the
knowledge of these transition points it is possible to construct the phase diagram
in the $(T/T_c,\;\rho/\rho_0)$ plane, Fig.~5. Here $T_c$ corresponds
to the critical point at $\rho/\rho_0=0.5$ where the usual spherical model
undergoes a second order transition working at zero magnetic field
$h$~\cite{singh}. In the spherical model, the transition for
$\rho/\rho_0\neq 0.5$ is of first order, owing to the discontinuity
in the curve $h$ vs $M$ at $T<T_c$. This means that when $h=0$, the system
can pass from a magnetization $+M$ to $-M$. But in the present model we
fix $M$ to a certain value (constraint (8)), so flips from $+M$ to $-M$
are not possible here. Therefore, one expects a behaviour for every $M$
similar to what happens at $M=0$. This is why we obtain the same
smooth behaviour at every value of $\rho/\rho_0$.

\section{Numerical simulations in the framework of a cellular model}

The present description is close to former numerical approaches (cellular 
model) which were introduced in order to study the thermodynamic 
properties and cluster content of an excited disordered system of
particles~\cite{elattari,boose}.
However in these simulations the system was chosen to be 
homogeneous ($\rho/\rho_0=1$) and the numerical procedure did not imply 
that the system was in thermodynamic equilibrium. There was no sign for the 
existence of a thermodynamic phase transition.

We present here the results of numerical simulations which generalise in 
the framework of the canonical ensemble the former calculations. We 
consider a finite volume of space $V$ in $3D$ made of unit cells of linear 
dimensions $d=1.8 \,\mathrm{fm}$, $V=N^3.d$. Each cell is either occupied or 
unoccupied, the total number of particles is $A$ such that the normalized 
density $\rho/\rho_0=A/N^3$. The particles interact by means of a
nuclear~\cite{wilets}  and Coulomb interaction.

For fixed density and temperature the equilibrium energy of the system is 
obtained by means of the following algorithm~\cite{heermann}. Assign an initial 
configuration with $A$ particles in $N^3$ cells. Select randomly with 
equal probability one of the two following procedures. In the first 
procedure one selects randomly a particle among the $A$ occupied cells and
chooses a new position of this particle inside its cell. This changes the 
energy of the system by an amount $\Delta E$ which can be positive or 
negative. At this stage one applies the Metropolis Monte Carlo algorithm
before returning to the procedure selection step.
In the second procedure one selects randomly a particle ``$a$'' in one of 
the occupied cells and an empty cell ``$b$'' among the $N^3-A$ empty cells.
One creates a particle at the center of the cell ``$b$'' with the same 
charge (proton or neutron) as the charge of the particle at ``$a$''. One 
destroys the particle at ``$a$'' and hence guarantees total mass and charge 
conservation. This changes the energy of the system by an amount $\Delta E$
and one applies again the Metropolis Monte Carlo algorithm before going
back to the procedure selection step. The average number of measurements is 
4000 corresponding to ten times more Metropolis steps.

Typical results are shown in Fig.~6. The fluctuation of the results is not 
larger than the size of the dots. The energy per particle in the abscissa 
is the sum of the interaction and the kinetic energy which gives the 
trivial contribution $3T/2$. We checked the consistency of the procedure by 
means of numerical simulations of the kinetic energy of a system of free 
particles from a Boltzmann distribution. The fluctuations in the kinetic 
energy are negligible.

The behaviour of the caloric curve is very similar to the curve obtained by 
means of the analytic model, see Fig.~4. Indeed, the line which goes 
through the points corresponds to a calculation with the spherical model and 
$V_0=5.5  \,\mathrm{MeV}$ for the attractive short range interaction. It is worthwile 
to notice that there is no sign for the existence of a phase transition. 
If however we interpret the results in the framework of the
spherical model, the observed smooth behaviour would correspond indeed to a
continuous transition reflected on the discontinuity in the derivative of
the specific heat at the temperature predicted by the model.

\section{Discussion and final remarks}

The fact that the description leads to a continuous phase transition can a priori 
seem somewhat surprising. Indeed, other models which are related to spin systems
like the lattice gas model clearly show a first order transition [4-7] which is not
reproduced here. One could argue that we have `softened' 
the strength of the transition by working in the framework of the
spherical model, because it corresponds to the limit $m\to\infty$ in the
$O(m)$ model. However, we remark that this is not the case. We obtain a
continuous transition owing to the constraint fixing the magnetization,
which prevents it from having the typical fluctuations which happen at the
first order transition points in these other models. In fact, one can 
consider this constraint in the Ising model and see numerically that 
the whole transition line corresponds to a second order phase
transition~\cite{isingpaper}. We have considered here the spherical model
approximation in order to obtain analytical expressions for the different
quantities and to incorporate short and long range interactions.
The controversy between the order of the transition
should work as an incentive
to experimental physics to look for the best way to identify
the specific properties of excited nuclei and nuclear matter and try to confirm
or infirm the existence of a discontinuous phase transition~[26-29].

The numerical simulations on the inhomogeneous model have been made in the framework
of the canonical ensemble, with strictly fixed volume and number of particles.
The fact that the results are in qualitative agreement with 
those obtained analytically in the framework
of the grand canonical ensemble indicates that volume and particle fluctuations
do not strongly affect the physical issue. 
It also shows that the interpretation of the continuum limit approximation
leading to the spherical model description makes sense.
Notice that in the experiment, volume and
exact particle number can hardly be rigorously fixed. It would nevertheless be
interesting to work out the dynamics in the framework of the microcanonical 
ensemble where the isolated nuclear system is treated rigorously~\cite{gross}. 
Whether this is possible  has to be investigated in practice.

Finally it would be of particular interest to work out averaged $n$-body correlation
functions which allows to determine the fragment size distributions. This can in 
principle be done by means of a 
straightforward extension of the partition function $Z$ to a
generating function by adding terms of the type 
$\exp\left[\sum_{i=1}^{N^3}(h_is_i+g_i(1-s_i))\right]$ 
into the sum over occupation numbers. 
Here $h_i$ and $g_i$ are arbitrary real constants and the probability of an
isolated cluster with $(i_1,\ldots, i_n)$ occupied cells can be
obtained by deriving this generating function with respect to specific parameters 
$h_i$ and $g_i$ (which fix the surface of the cluster) 
and putting the parameters to zero. 
The determination of the cluster formation probabilities would then 
lead to a complete description of the fragmentation
of the considered system in the $(\epsilon,\;\rho,\;\beta)$ space.

\newpage

\begin{figure}

{\large \bf Figure caption}\\

\caption{Specific heat calculated from (13) with $\rho/\rho_0=0.90$,
$V_0=8  \,\mathrm{MeV}$, and different sizes of the system $(N=10,20,\ldots,50)$.}

\caption{Caloric curve obtained with an attractive short range
(nearest-neighbour) interaction with strength $V_0=8  \,\mathrm{MeV}$, 
$A=N^3\rho/\rho_0=250$.}

\caption{Caloric curve obtained with an attractive long range
interaction $(n=20)$ with strength $W_0=8  \,\mathrm{MeV}$.}
              
\caption{Caloric curve for systems which experience both an attractive
short range (nearest-neighbour) with strength $V_0=8  \,\mathrm{MeV}$ and a repulsive
long range interaction with strengths given by (15) $(n=20)$. Here 
$Z_p/A=0.394$ and $d=1.8  \,\mathrm{fm}$, see eq.(14).}

\caption{Phase diagram $T/T_c$ vs. $\rho/\rho_0$. The transition
line is read from the behaviour of the specific heat. The point $T/T_c=1$
corresponds to the continuous phase transition in the thermodynamic limit
of the usual spherical model.}

\caption{Caloric curve obtained by means of Metropolis Monte Carlo 
simulations. The points correspond to numerical results for $N^3 = 216$,
$\rho/\rho_0 = 0.5$, $A = 108$ and total charge number $Z_p = 47$. The full
line corresponds to the results obtained through the spherical model with
$V_0 = 5.5  \,\mathrm{MeV}$. The arrow labelled $T_0$ indicates the phase
transition temperature. See explanations and comments in the text.}

\end{figure}

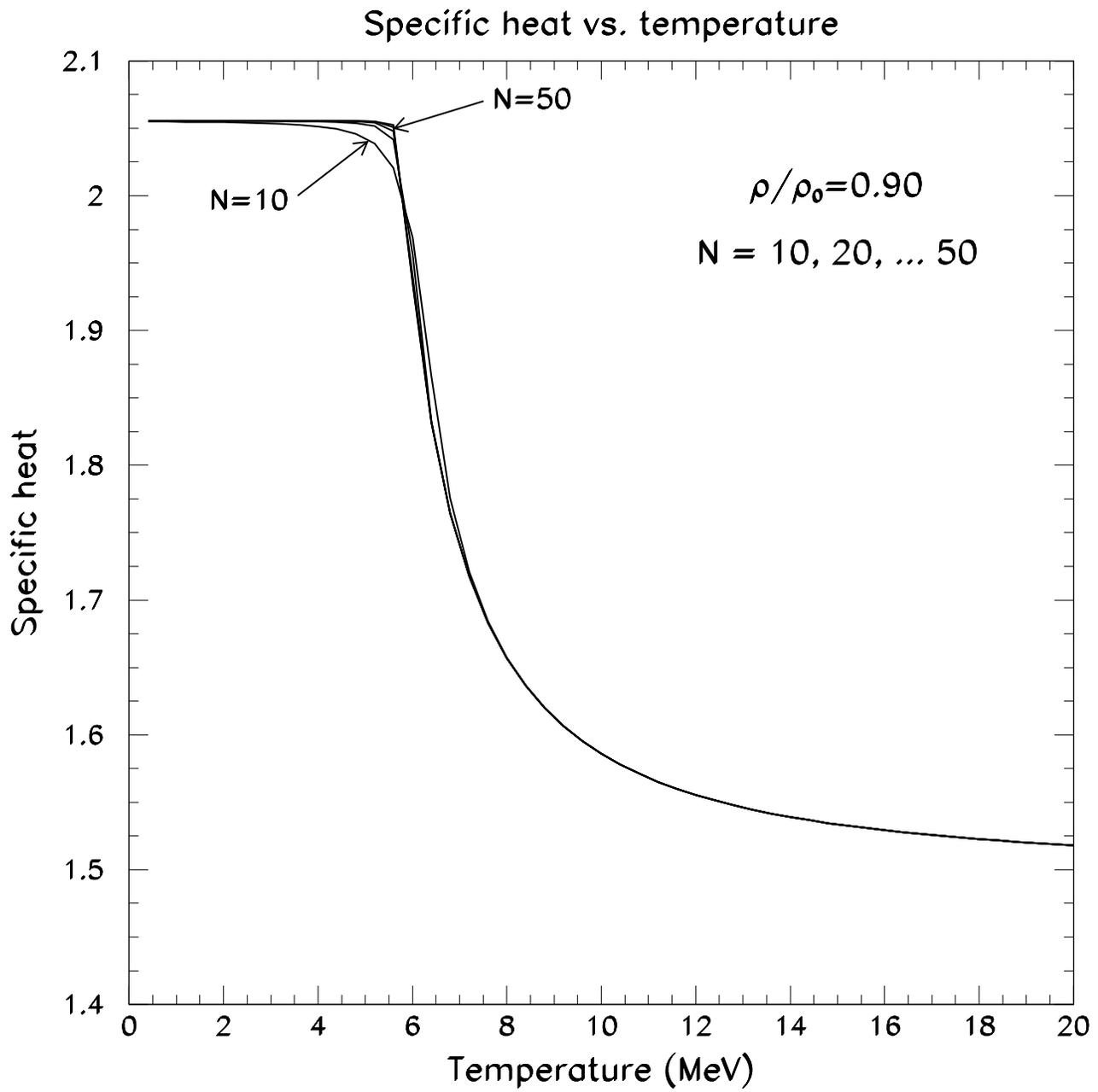
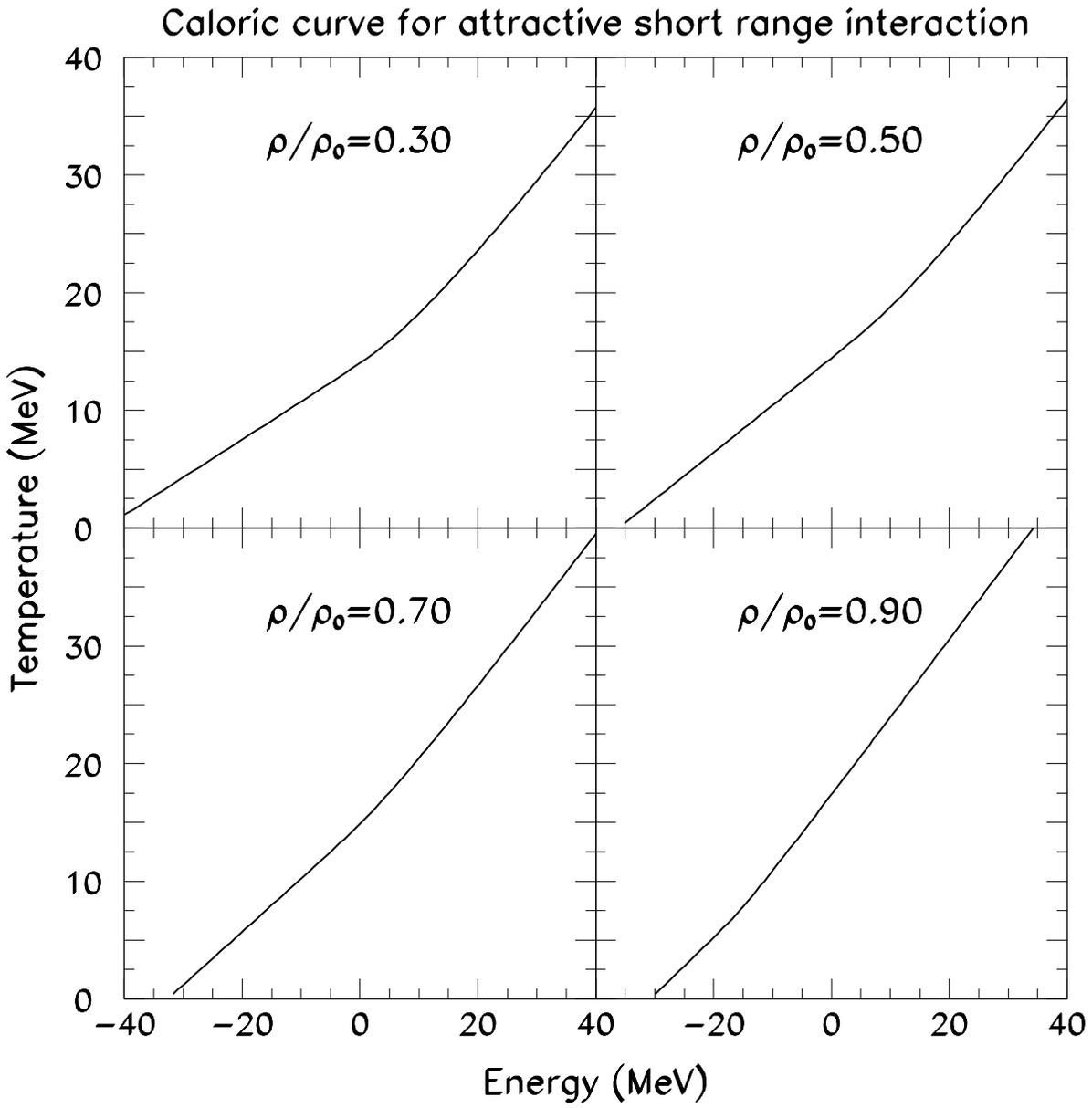
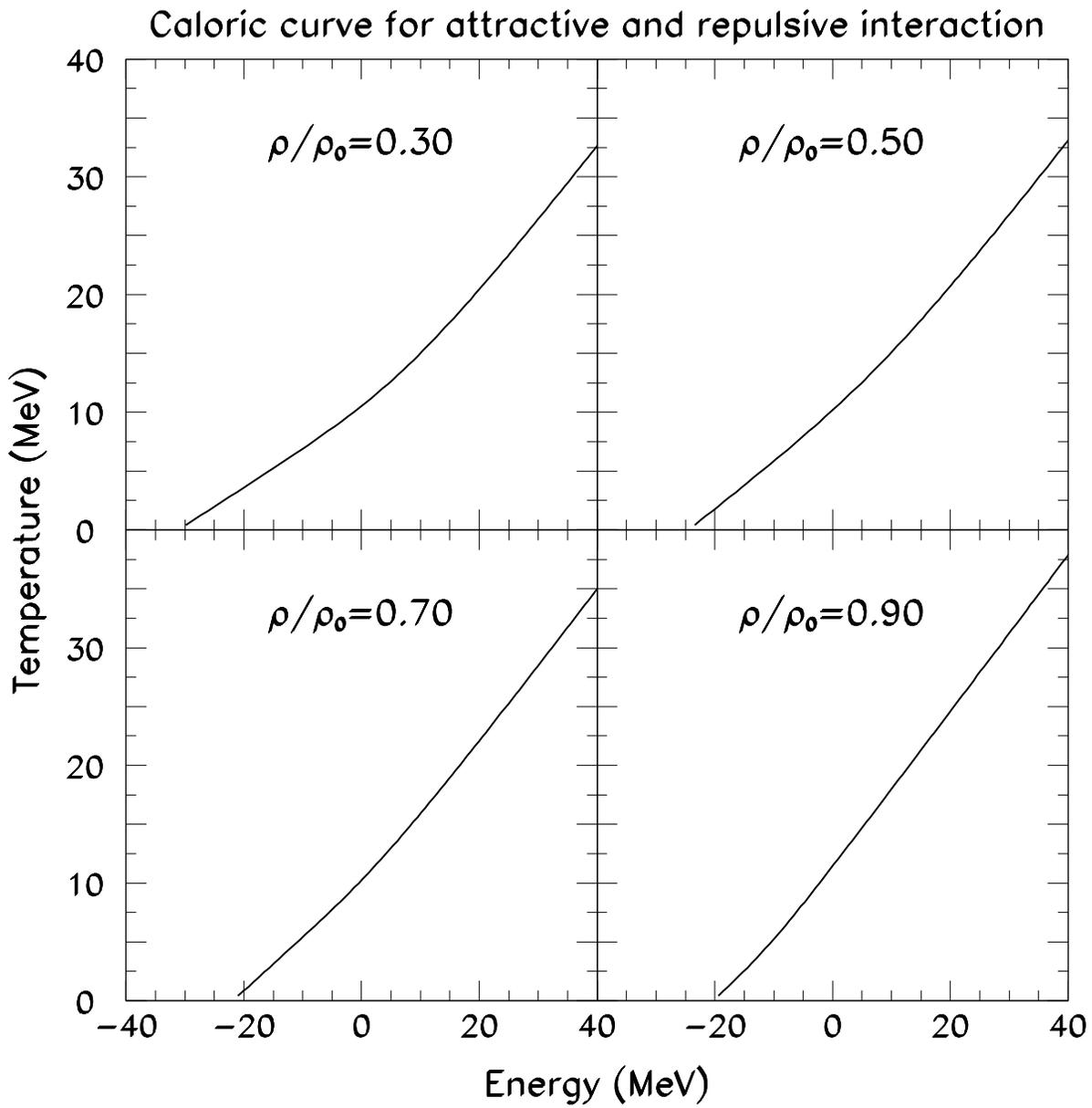
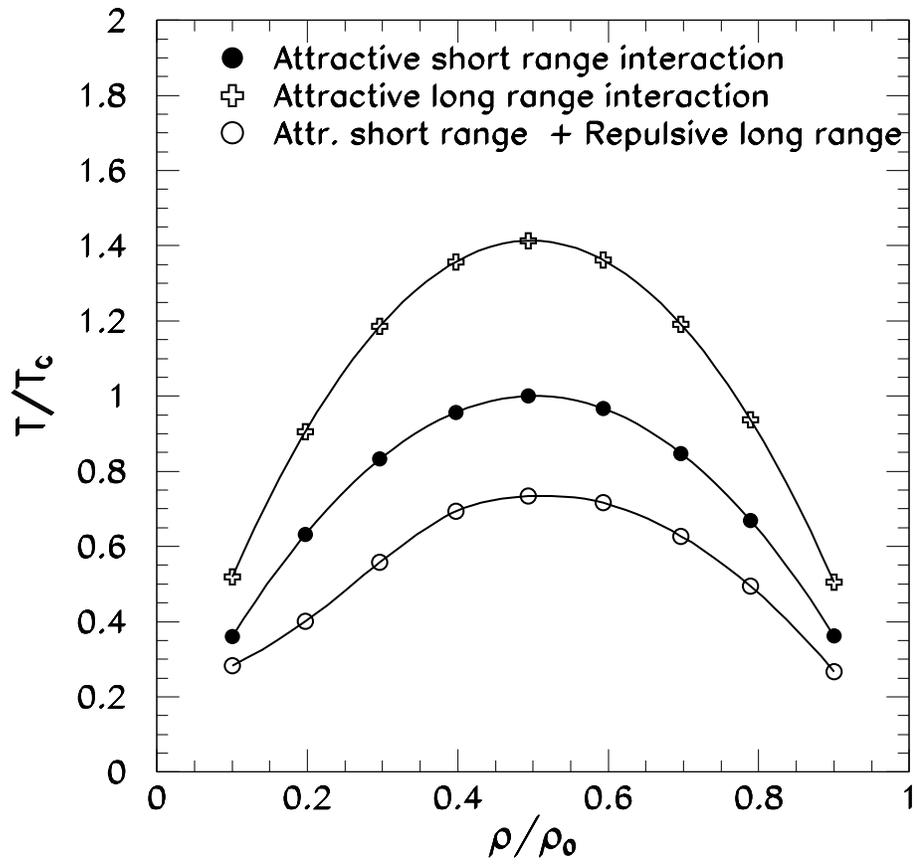
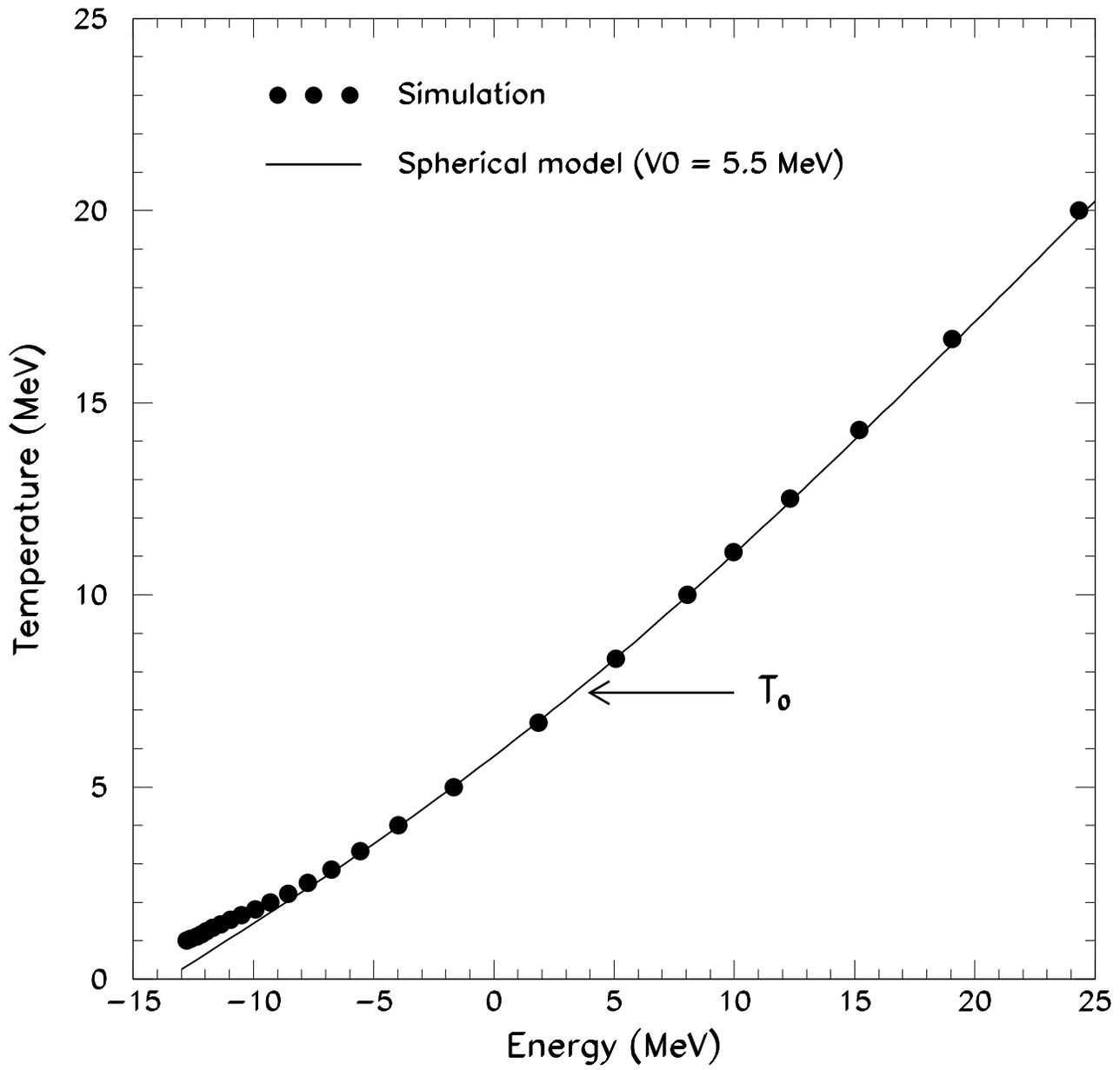
\end{document}